\documentclass[twocolumn,journal,10pt]{IEEEtran}
\usepackage{epsfig,amsfonts,subfig,psfrag}
\usepackage[nolist]{acronym}
\usepackage{graphicx,cite,amssymb,amsmath,color}
\usepackage{psfrag}
\usepackage{lipsum}
\usepackage[dvipsnames]{xcolor}
\usepackage{tikz,pgfplots,relsize}
\usetikzlibrary{fadings}
\usetikzlibrary{shadows.blur}
\usetikzlibrary{shapes,arrows}
\usetikzlibrary{calc}
\usetikzlibrary{patterns}
\begin{document}
\begin{acronym}
\acro{CRDSA}{contention resolution diversity slotted ALOHA}
\acro{CSA}{coded slotted ALOHA}
\acro{IRSA}{irregular repetition slotted ALOHA}
\acro{IoT}{internet of things}
\acro{LDPC}{low density parity check}
\acro{MAC}{medium access control}
\acro{NOMA}{non-orthogonal multiple access}
\acro{RA}{random access}
\acro{SA}{slotted ALOHA}
\acro{SIC}{successive interference cancellation}
\acro{TDMA}{time division multiple access}
\end{acronym}

\title{\LARGE From 5G to 6G: Has the Time for Modern Random Access Come?}
\author{Federico Clazzer, Andrea Munari, Gianluigi Liva, Francisco Lazaro, Cedomir Stefanovic, Petar Popovski
\thanks{
F. Clazzer, A. Munari, G. Liva and F. Lazaro are with Institute of Communications and Navigation of the Deutsches Zentrum fur Luft- und Raumfahrt (DLR), 82234 Wessling, Germany (e-mail: \{Federico.Clazzer, Andrea.Munari, Gianluigi.Liva, Francisco.LazaroBlasco\}@dlr.de). C. Stefanovic and P. Popovski are with Department of Communication Systems, Aalborg University, 9220 Aalborg, Denmark (e-mail: \{cs, petarp\}@es.aau.dk.)}} \maketitle
\date{\today}
\thispagestyle{empty} \setcounter{page}{0}


\begin{abstract}
This short paper proposes the use of modern random access for IoT applications in 6G. A short overview of recent advances in uncoordinated medium access is provided, highlighting the gains that can be achieved by leveraging smart protocol design intertwined with advanced signal processing techniques at the receiver. The authors' vision on the benefits such schemes can yield for beyond-5G systems is presented, with the aim to trigger further discussion.
\end{abstract}

{\pagestyle{plain} \pagenumbering{arabic}}


\section{Introduction}

\IEEEPARstart{T}{he} growing success of the \ac{IoT} systems demands for wireless technologies that are capable of attaining reliable data transmission with high energy- and spectral-efficiency~\cite{Boccardi14:MAG}.
The task is particularly challenging due to the peculiarities of \ac{IoT} traffic. In fact, \ac{IoT} transceivers often operate by generating short data packets in a sporadic, and sometimes unpredictable manner. When the network consists of a massive number of nodes -- as foreseen by the \ac{IoT} paradigm -- the sporadic transmission of short data packets makes it difficult to devise an efficient resource allocation policy.
This issue can be addressed by employing \ac{RA} protocols: the network nodes access the communication medium in an uncoordinated fashion, without the need of establishing a costly (in terms of wireless resources) negotiation with the resource manager.

Most \ac{RA} solutions used in modern cellular systems, such as 5G, SigFox~\cite{SigFox}, and LoRaWAN~\cite{LoRa}, are conceptually very similar to the first \ac{RA} protocols introduced in the 1970s, in particular to the family of ALOHA-like schemes \cite{Abramson1970}. However, when employed in a heavily-overloaded network granting access to a massive number of terminals, these basic protocols yield a throughput performance that is largely suboptimal. Moreover, the current standardization outlook suggests that 5G access will be connection-oriented~\cite{TS38.321}, with \ac{RA} adopted solely for the connection establishment. This approach is known to be highly inefficient for \ac{IoT} services, e.g. smart grid, industry $4.0$, V2X communications, etc.~\cite{7397849}, urging the investigation of grant-free access methods, 
where the IoT devices deliver data through random access.

A recent flurry of research activities has enlightened drastic improvements for \ac{RA}, achieved thanks to proper protocol design and to the use of sophisticated receiver algorithms, see e.g. \cite{Paolini15:MAG} and references therein. An increase of several orders of magnitude for the sustainable channel load can be granted, dramatically boosting the efficiency of the underlying communication systems.
These developments, pioneered mostly in the context of satellite machine-type communications \cite{Casini2007,Liva2011,Clazzer18:ECRA,delRioHerrero_2012}, have led to the establishment of \emph{modern \ac{RA} protocols}.
This new paradigm \cite{Berioli2016} relies on a design that is tailored to harness information from multi-user interference, and on a tight integration of physical and \ac{MAC} layers functionalities at the receiver side.
Ten years after the introduction of the first modern \ac{RA} protocols, already a few commercial satellite communication systems, e.g. DVB-RCS2 \cite{dvbrcs2}, benefit from their remarkable performance, showcasing their maturity. Their undeniable benefits pave the way to their adoption in the next generation terrestrial cellular networks. Based on these considerations, this paper reviews the basic principles of \emph{modern \ac{RA}} algorithms and highlights potential applications to -- and synergies with -- 6G systems.

\section{Modern Random Access Protocols}

Recent years witnessed a renewed research and industry interest towards the development of efficient uncoordinated \ac{MAC} schemes.
The quest climaxed in the design of ALOHA-based solutions with performance competitive to that of their coordinated counterparts in terms of throughput and reliability, paving the road for the application of \emph{modern \ac{RA} protocols} to many \ac{IoT} scenarios of 6G.
The key idea to achieve such a remarkable result is to constructively embrace interference, allowing transmitters to send multiple copies of their packets, and performing \ac{SIC} -- possibly combined with other advanced signal processing techniques -- at the receiver side, to recover information.

\begin{figure*}
\centering
 \subfloat[Example of operation for a \emph{modern \ac{RA} protocol}.]
   {\includegraphics[width=7cm]{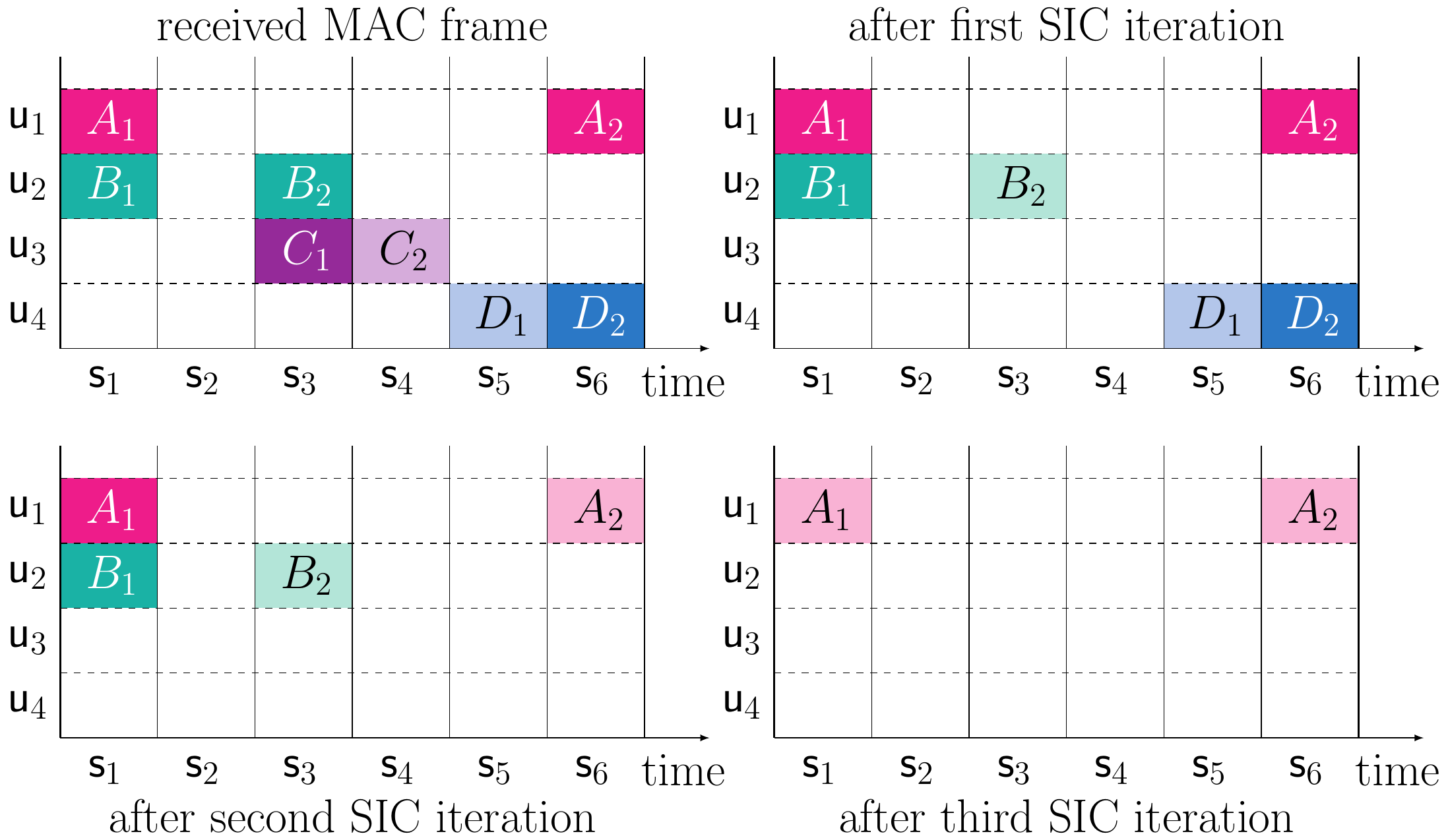}
   \label{fig:crdsa_SIC}}
 \hspace{2mm}
 \subfloat[Throughput vs. channel load.]
   {\includegraphics[width=4.8cm]{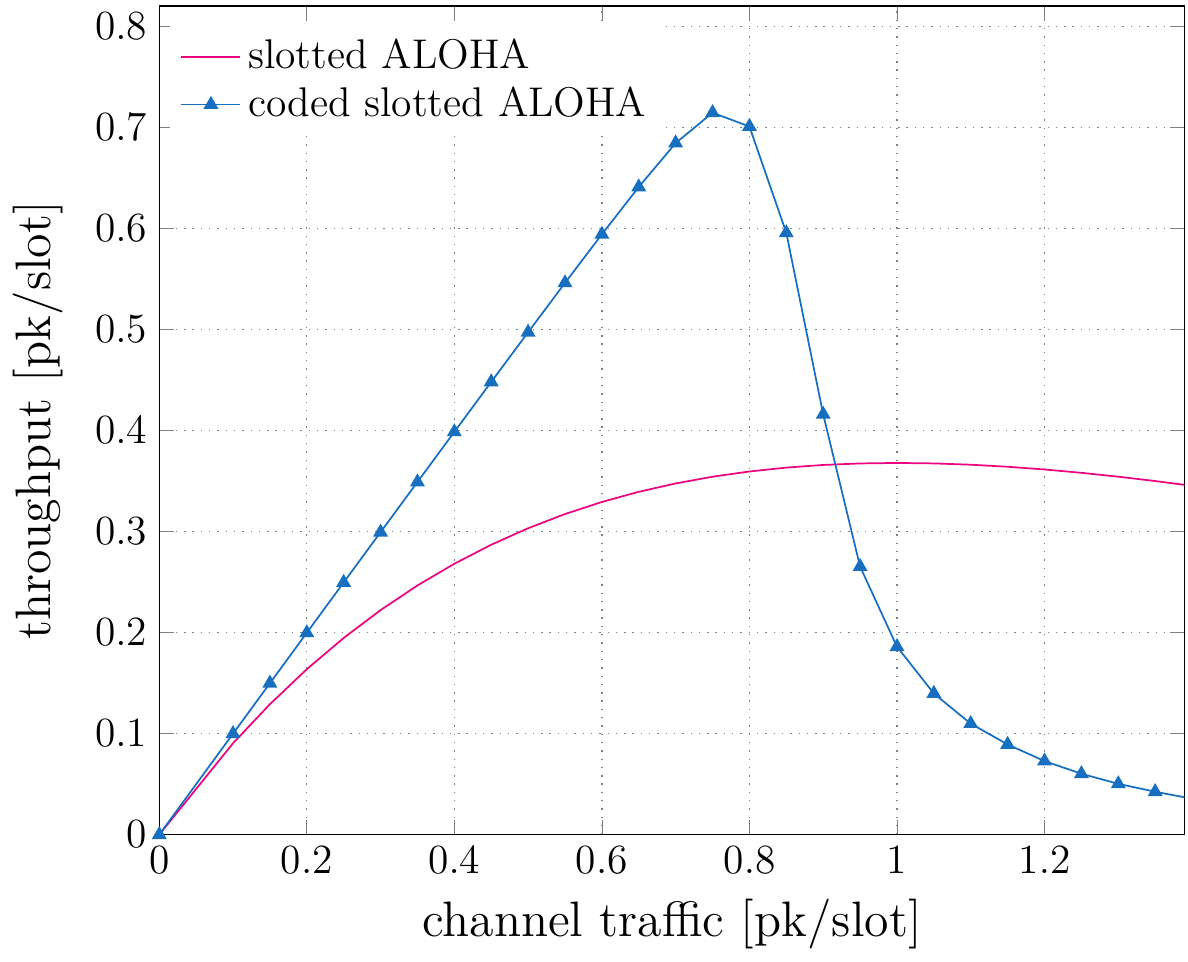}
   \label{fig:tru}}
   \hspace{2mm}
 \subfloat[Throughput at target packet loss rate.]
   {\includegraphics[width=4.8cm]{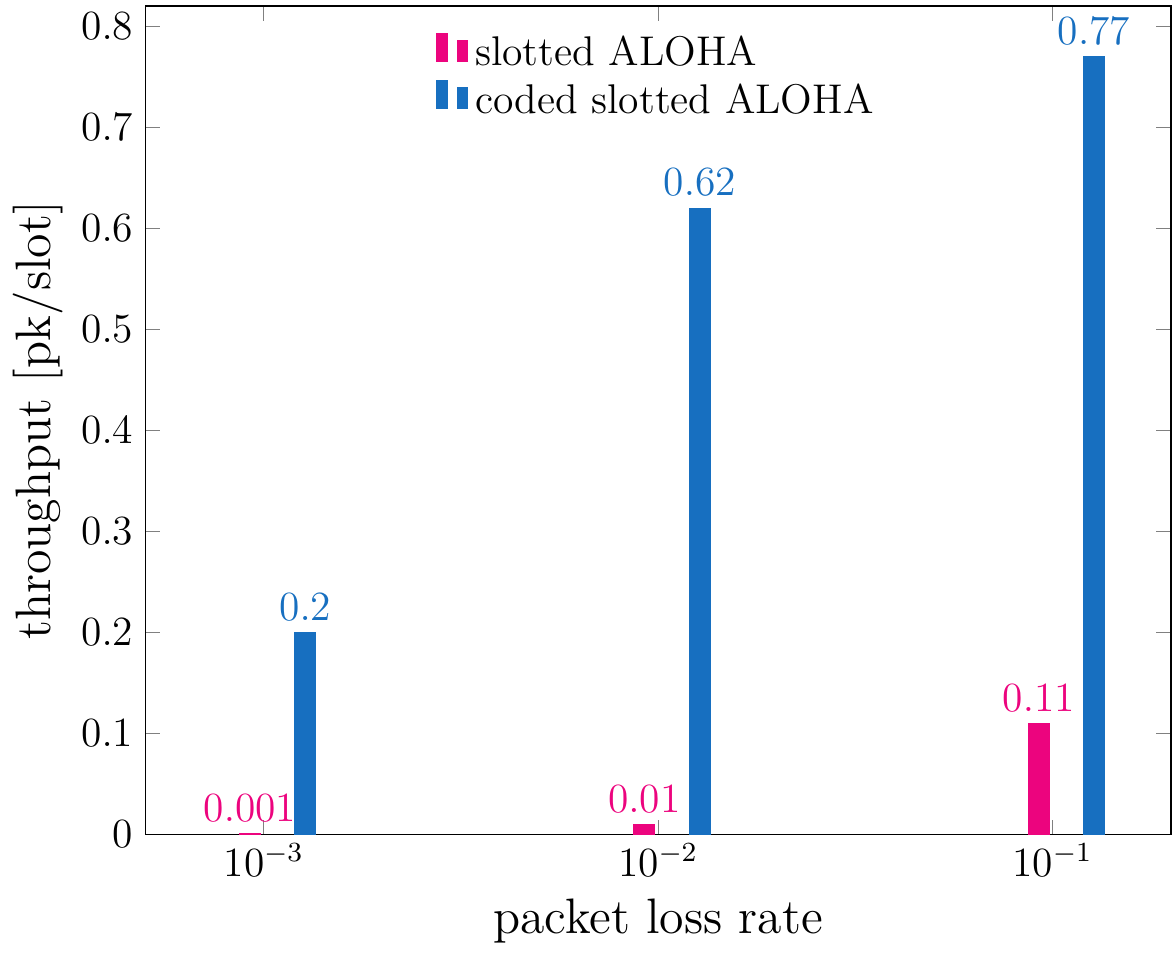}
   \label{fig:barplot}}
 \caption{Operation and performance of \emph{modern \ac{RA}} protocols.}
 \label{fig:plot_all}
\end{figure*}

This principle was first applied in \ac{CRDSA} \cite{Casini2007}, whose operation is shown in Fig.~\ref{fig:crdsa_SIC}.
In this example, each of four users transmits two copies of its data packet, randomly placing them over a \ac{MAC} frame of predefined duration. In turn, the receiver buffers the whole frame, and starts decoding by looking for interference-free, i.e. decodable, packets.
In the example, this happens for the second copy of the data unit of user $\mathsf u_3$ (denoted as $C_2$) and the first copy sent by user $\mathsf u_4$ (denoted as $D_1$).
The interference generated by the other copies ($C_1$ and $D_2$) of the retrieved packets can then be subtracted from the signal received over the respective slots, potentially rendering other data units decodable.
The interference cancellation process is then iterated, making it possible to retrieve all transmitted information.\footnote{We note that the approach of \emph{modern \ac{RA} protocols} can be seen as a variant of \ac{NOMA}. However, \ac{NOMA} solutions focus mainly on the downlink, and as such do not address randomness in user activity, often encountered in \ac{IoT}.}

The full potential of this idea can be unleashed by allowing users to send a number of packets that follows an optimized distribution, and by applying smart forms of coding across transmitted data units, as proposed by the family of \ac{CSA} protocols  \cite{Liva2011}, \cite{Paolini2014}.
Following this approach, \emph{modern \ac{RA}} schemes more than double the throughput of basic ALOHA for moderate frame size (Fig.~\ref{fig:tru}), and have been shown to asymptotically approach the $1\, \rm{pk/slot}$ performance achieved by orthogonal access (e.g. TDMA), for sufficiently long frames \cite{Narayanan2012}.
Even more interestingly, for quality of service-constrained applications, \ac{CSA} protocols can be tuned to serve a much larger number of users, i.e. a higher load, than traditional \ac{RA} solutions, offering orders-of-magnitude improvements in reliability (Fig.~\ref{fig:barplot}).
Such results, combined with the observation that they can be obtained with very little additional complexity at the transmitter, make \emph{modern \ac{RA}} especially appealing for \ac{IoT} scenarios in 6G.

Starting from these remarks, a number of promising schemes have recently been proposed, combining diversity ALOHA with \ac{SIC} in asynchronous \cite{Clazzer18:ECRA} or spread-spectrum setups \cite{delRioHerrero_2012}, as well as exploring complementary ideas such as operation without frames of predefined duration \cite{Stefanovic2013_J}, or the use of multiple receivers \cite{Munari2013}.

\section{Vision for Random Access in 6G-IoT}

While it is true that alluring gains can be obtained with \emph{modern random access protocols}, further study is necessary to understand their true potential in 6G. Indeed, although such schemes have already been successfully adopted in satellite communications, their application to terrestrial scenarios poses novel challenges. On the one hand, understanding the synergies of \emph{modern \ac{RA}} with techniques such as massive MIMO, OFDM, \ac{NOMA}, or sparse signal processing, triggers the opportunity for optimized system design tailored to 6G~\cite{deCarvalho2017}.
On the other hand, specific channel and traffic characteristics have to be taken into account. From this standpoint, the classic dichotomy between pure \ac{RA} -- no information on transmitters is available -- and scheduled access -- perfect knowledge is available -- no longer applies to several \ac{IoT} settings. The use of data-driven methods lends itself to the design of new generalized \ac{RA} protocols, where the receiver has a certain side information about the (possibly correlated) activation patterns of the devices.

In light of their features and performance, the 6G ecosystem naturally calls for \emph{modern \ac{RA}} schemes as key enablers for \ac{IoT} use-cases.

\bibliographystyle{IEEEtran}
\bibliography{IEEEabrv,references}
\end{document}